\documentclass{amsart}
\usepackage{hyperref}
\usepackage{graphicx}
\usepackage{curves}
\usepackage{enumerate}
\newcommand*{\mailto}[1]{\href{mailto:#1}{\nolinkurl{#1}}}

\newcommand{\R}{\mathbb{R}}
\newcommand{\Z}{\mathbb{Z}}

\newcommand{\C}{\mathbb{C}}

\newcommand{\M}{\mathbb{M}}

\newcommand{\be}{\begin{equation}}
\newcommand{\ee}{\end{equation}}
\newcommand{\bea}{\begin{eqnarray}}
\newcommand{\eea}{\end{eqnarray}}

\newcommand{\ol}{\overline}
\newcommand{\pa}{\partial}

\newcommand{\I}{\mathrm{i}}
\newcommand{\E}{\mathrm{e}}

\newcommand{\re}{\mathop{\mathrm{Re}}}

\DeclareMathOperator{\res}{Res}

\newcommand{\noprint}[1]{}

\newcommand{\si}{\sigma}
\newcommand{\la}{\lambda}

\numberwithin{equation}{section}

\newcommand{\sigI}{\begin{pmatrix} 0 & 1 \\ 1 & 0 \end{pmatrix}}

\begin{document}
	\title[Wave phenomena of the Toda lattice]{Wave phenomena of the Toda lattice with steplike initial data}
\author[J. Michor]{Johanna Michor}
\address{Faculty of Mathematics\\ University of Vienna\\
Oskar-Morgenstern-Platz 1\\ 1090 Wien\\ Austria\\ and International Erwin Schr\"odinger
Institute for Mathematical Physics\\ Boltzmanngasse 9\\ 1090 Wien\\ Austria}
\email{\href{mailto:Johanna.Michor@univie.ac.at}{Johanna.Michor@univie.ac.at}}
\urladdr{\href{http://www.mat.univie.ac.at/~jmichor/}{http://www.mat.univie.ac.at/\string~jmichor/}}

\thanks{Research supported by the Austrian Science Fund (FWF) under Grant No.\ V120.}
\thanks{Phys.\ Lett.\ A 380, 1110--1116 (2016)}

\begin{abstract}
We give a survey of the long-time asymptotics for the Toda lattice with steplike constant 
initial data using the nonlinear 
steepest descent analysis and its extension based on a suitably chosen $g$-function.
Analytic formulas for the leading term of the asymptotic solutions of the Toda shock and rare\-faction problems (including the case of overlapping background spectra) are given and complemented by numerical simulations. 
We provide an explicit formula for the modulated solution in terms of Abelian integrals on the underlying hyperelliptic Riemann surface.  
\end{abstract}

\maketitle

\section{Introduction}

We are interested in the long-time behavior of an infinite particle chain with
nonlinear nearest neighbor interactions when the chain is subjected to shock or rarefaction type
initial conditions.
The continuous spectrum of the underlying Lax operator consists of two intervals which might overlap  
and their mutual location produces essentially different types of asymptotic solutions. 
These wave phenomena were first discovered numerically in \cite{hs, hfm}, 
a rigorous investigation of the limiting behavior as $t\to \infty$ has been carried out so 
far only for special initial values \cite{dkkz, Kb, vdo}. 
This introductory article gives an overview of the new set of results on the Toda shock and rarefaction problems,
in particular, we present the leading term of the long-time asymptotic solution for arbitrary steplike constant
initial data. The mathematical proof is given in \cite{emt14} and a forthcoming paper.

\begin{figure}[ht]
\centering
\includegraphics[width=6cm]{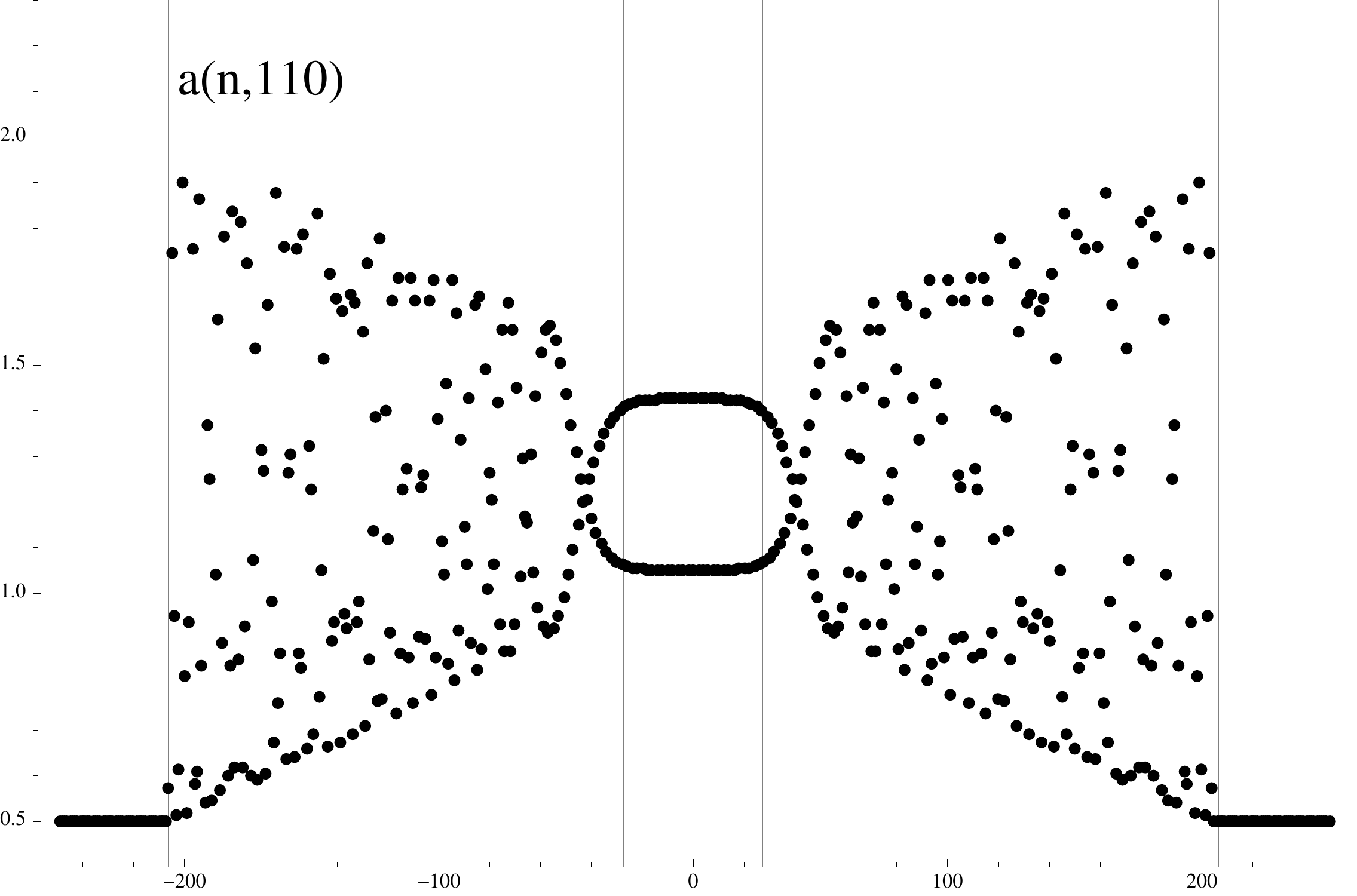}
\hfill
\includegraphics[width=6cm]{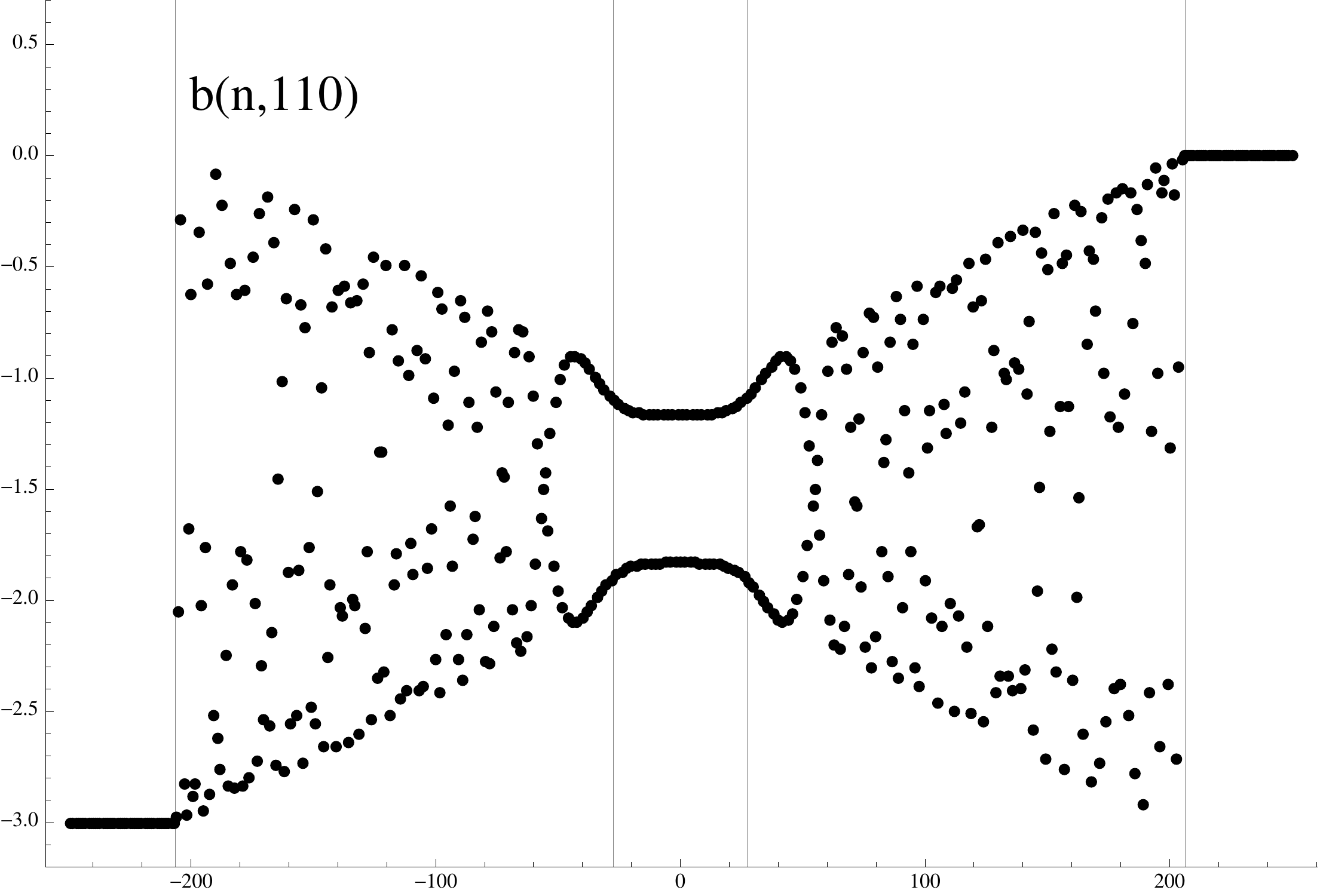}
\caption{{\small Toda shock problem at $t=110$ with non-overlapping background spectra of equal length, 
$\sigma(H_{\ell})=[-4, -2]$ and  $\sigma(H_r)=[-1,1]$. % ($a_{\ell}=\frac{1}{2}$, $b_{\ell}=-3$).
}} \label{shocknools}
\end{figure}

Consider the doubly infinite Toda lattice (\cite{tjac, toda}) in Flaschka's variables
\begin{align} \label{tl}
	\begin{split}
\dot b(n,t) &= 2(a(n,t)^2 -a(n-1,t)^2),\\
\dot a(n,t) &= a(n,t) (b(n+1,t) -b(n,t)),
\end{split}
\end{align}
$(n,t) \in \Z \times \R$, where the dot denotes differentiation with respect to time. We study a
steplike initial profile
\begin{align} \label{ini1}
\begin{split}	
& a(n,0)\to a_{\ell}, \quad b(n,0) \to b_{\ell}, \quad \mbox{as $n \to -\infty$}, \\
& a(n,0)\to a_r \quad b(n,0) \to b_r, \quad \mbox{as $n \to +\infty$},
\end{split}
\end{align}
where the left and right background Jacobi operators with constant coefficients 
$a_{\ell,r} > 0$, $b_{\ell,r} \in \R$,  
$$
(H_{\ell,r} y)(n)=a_{\ell,r} y(n-1)+b_{\ell,r} y(n)+a_{\ell,r} y(n+1), \quad n \in \Z,
$$
have spectra in the following general location: 
$$
\si(H_{\ell}) \neq \si(H_r).
$$
Each of these spectra consist of one interval and in the literature so far, only background
spectra of equal length (that is, when $a_{\ell}=a_r$) have been partly investigated. 
Under this assumption, there are two classical cases, distinguished by the conditions
$\inf \sigma(H_{\ell}) < \inf \sigma(H_r)$ (the Toda shock problem) and  
$\inf \sigma(H_{\ell}) > \inf \sigma(H_r)$ (the Toda rarefaction problem). 
The Toda shock problem with non-overlapping background spectra was studied 
by Venakides, Deift, and Oba \cite{vdo}. This case is depicted in Fig.~\ref{shocknools}, where
the numerically computed solution corresponding to the step $b(n,0)= 0$ if $n\geq 0$, $b(n,0)= -3$ if $n< 0$, and
$a(n,0) \equiv \frac{1}{2}$ is plotted at a frozen time $t=110$ for $500$ plotpoints around the origin. 
These data correspond to a pure step without solitons. In areas where the functions $n \mapsto a(n,t)$ 
and $n \mapsto b(n,t)$ seem to be continuous this is due to scaling, since we have plotted a large number 
of particles, and also due to the $2$-periodicity in space. So one can think of the two lines in 
the middle region as the even- and odd-numbered particles of the lattice.
There are five principal regions in the half plane $n/t$ divided by rays $\pm \xi_{cr}$, $\pm \xi_{cr}^\prime$,
$\xi_{cr}^\prime < \xi_{cr}$,
with transitional regions around the rays. The points $n_{cr}=\xi_{cr} \times 110$ are plotted as vertical lines 
to mark the different regions.
In the middle region $|\frac{n}{t}| < \xi_{cr}^\prime$, 
the solution can be asymptotically described by a periodic Toda solution of period two, 
which was the main result of \cite{vdo}. 
In the region $\frac{n}{t} > \xi_{cr}$, the solution is asymptotically close to the constant right background solution 
$(a_r, b_r)$, and in the domain $\frac{n}{t} < -\xi_{cr}$, it is close to the left background $(a_{\ell}, b_{\ell})$.
For the remaining region $ \xi_{cr}^\prime < |\frac{n}{t}| < \xi_{cr}$, it was conjectured in \cite{vdo}
that the solution is asymptotically close to a modulated single-phase quasi-periodic solution,
but despite some follow-up publications (\cite{bk, bk2, Kb}), this problem remained open. 
We give a precise analytical description of this solution in Sec.~\ref{shockno} and \cite{emt14}.  
For the transitional regions around $\pm\xi_{cr}$, one can expect the appearance 
of asymptotic solitons (compare \cite{bdmek}), whereas the transitional regions around $\pm\xi_{cr}^\prime$ have not
been studied yet from an analytical point of view. Soliton asymptotics in the region 
$|\frac{n}{t}| > \xi_{cr}$ have been described in \cite{bdme} using the inverse scattering transform.

\begin{figure}[ht]
\centering
\includegraphics[width=5cm]{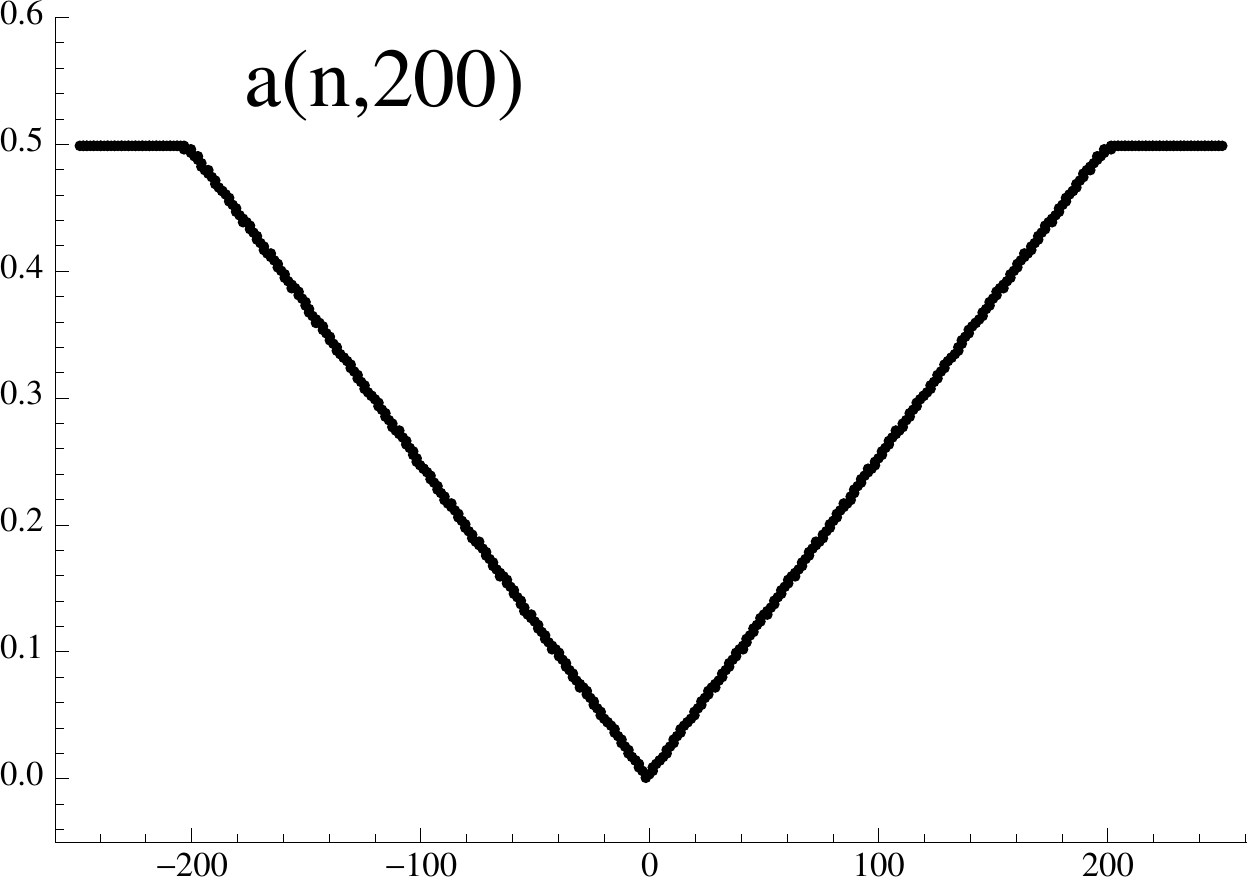}
\hspace{1cm}
\includegraphics[width=5cm]{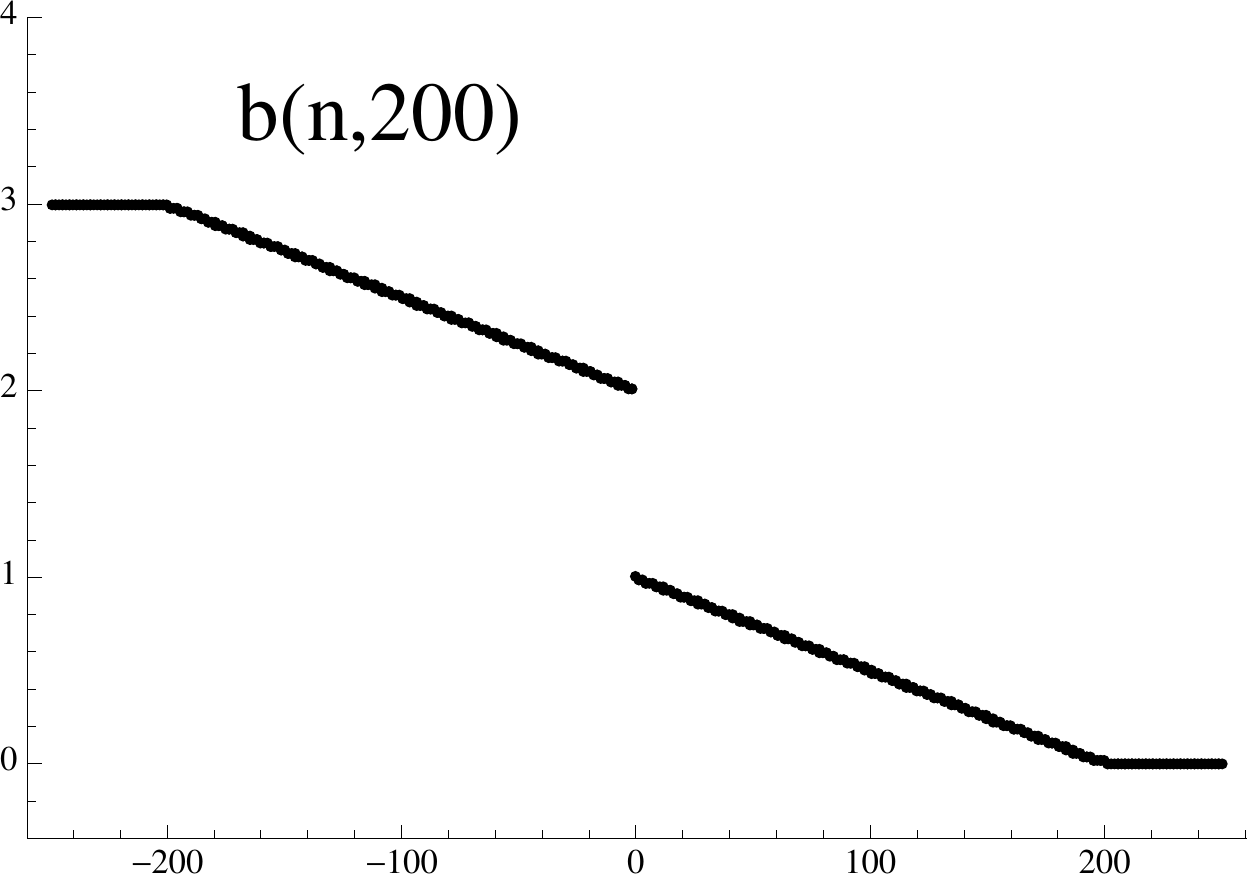}
\caption{{\small Toda rarefaction problem with non-overlapping background spectra $\sigma(H_{\ell})=[2, 4]$ and  $\sigma(H_r)=[-1,1]$.  \label{rarenools} }}% ($a_{\ell}=0.4$, $b_{\ell}=2$).}} 
\end{figure}

As for the Toda rarefaction problem, the only known result is by Deift, Kamvissis, Kriecherbauer, and 
Zhou \cite{dkkz}, who considered non-overlapping background spectra in the case $t \to \infty$ with $n$ fixed, 
which corresponds to the transitional region around $0$ in Fig.~\ref{rarenools}. 
The other regions have not been studied until now, and we present the leading term of 
the asymptotic solutions in Sec.~\ref{ss:no}. 
The proof is deferred to a forthcoming paper.

In this review we allow any possible location for the intervals $\sigma(H_{\ell})$, $\sigma(H_r)$. 
First of all, we extend the notion of Toda shock respectively rarefaction to 
include background spectra of different length, $a_{\ell} \neq a_r$. If they overlap, they have to satisfy 
(in addition to the conditions on the infima of the spectra) the condition 
$\sup \sigma(H_{\ell}) \leq \sup \sigma(H_r)$ for shock, or  
$\sup \sigma(H_r) \leq \sup \sigma(H_{\ell})$ for rarefaction, respectively. 
Otherwise, if one background spectrum is embedded in the other,
$\sigma(H_{\ell}) \subset \sigma(H_r)$ or $\sigma(H_r) \subset  \sigma(H_{\ell})$,  
it will produce mixed cases with a region where the solution is asymptotically close to a modulated two band Toda lattice 
solution and a second region where the asymptotic solution is given by a slope as in Fig.~\ref{rarenools}. These mixed cases are described in Sec.~\ref{mixed}. 
Let us mention that the limiting behavior for a flat background $\sigma(H_{\ell})=\sigma(H_r)$ 
is well understood by now, see the review article 
\cite{KTb} for decaying and \cite{KT} for quasi-periodic background operators.

\section{The Riemann--Hilbert problem for steplike constant background}

Without loss of generality we choose $a_r=\frac{1}{2}$, $b_r=0$ as the right initial data 
by shifting and scaling the spectral parameter $\la$ in the isospectral problem $H(t) y= \la y$. 
Here $H(t)$ is the Jacobi operator associated with the coefficients $a(t), b(t)$.
Suppose that the initial data \eqref{ini1} decay to their backgrounds exponentially fast (in the sense of \cite{emt14}).
Denote the spectra of the background operators by 
$$
I_{\ell} = \sigma(H_{\ell})=[b_{\ell} - 2a_{\ell}, b_{\ell} + 2a_{\ell}], \quad I_r= \sigma(H_r)=[-1,1].
$$  
The (absolutely) continuous spectrum of $H(t)$ consists of a part 
$(I_{\ell}\cup I_r)\setminus (I_{\ell}\cap I_r) $ of multiplicity one and  
a part $I_{\ell}\cap I_r$ of multiplicity two (if present).  
We assume for simplicity that $H$ has no eigenvalues, so no solitons are present. 

The perturbed solution $(a,b)$ of \eqref{tl}, \eqref{ini1} can be computed via the inverse scattering transform (IST).
The case without step ($H_{\ell}=H_r$) is well-known (see \cite{tjac, toda}); the general steplike case 
applicable here has been analyzed in \cite{emtstp2, emt3}. 
To obtain the long-time asymptotics of this solution we use a modification of IST in the form of 
a Riemann--Hilbert problem (RHP) on the underlying Riemann surface formed by combining both background spectra. 
For example, for
non-overlapping background spectra the Riemann surface associated with the square root
\be \label{rs}
P(\la)=- \sqrt{(\la^2-1)((\la-b_{\ell})^2-4a_{\ell})}
\ee
is $2$-sheeted over the complex plane where one changes sheets along the segments $I_{\ell}$ and $I_r$.
The square root plays the role of the projection onto the complex plane.
A point on the Riemann surface is denoted by $p=(\la, \pm)$, $\la \in \C$, with $p=(\infty, \pm)=\infty_\pm$. 
Let $\Sigma_{\ell}$ and $\Sigma_r$ be clockwise oriented contours around the cuts $I_{\ell}$ and $I_r$ 
on the upper sheet of the Riemann surface.
The asymptotic solution can be read off from a vector-valued function $m$ defined on the upper sheet using 
the right scattering data,
$$
m(p,n,t) =
\begin{pmatrix} T(p,t) \psi_{\ell}(p,n,t) z^n(p),  & \psi(p,n,t)  z^{-n}(p) \end{pmatrix}.
$$
Here $T(p,t)$ is the right transmission coefficient and $\psi$, $\psi_{\ell}$ are the Jost solutions 
of $H(t)y=\la y$ which asymptotically look like the free solutions of the background operators $H_r$ and $H_{\ell}$. 
The function $z(p) = \la - \sqrt{\la^2 - 1}$, $|z(p)|<1$, is the Joukovski transform of the spectral parameter $\la$.
We extend $m$ to the lower sheet by the symmetry condition
$$
m(p^*) = m(p) \sigI,
$$
where $p^*=(\la,-)$ is the flip image of $p=(\la,+)$ on the lower sheet.
With this extension, the scattering relations between the Jost solutions translate to jump conditions for $m$ 
along $\Sigma_{\ell}$ and $\Sigma_r$. To formulate them, let $m_+(p)$ denote the limit
of $m(\zeta)$ as $\zeta \to p$ from the positive side of $\Sigma$; the positive side is the one which
lies to the left of $\Sigma$ as one traverses $\Sigma$ in the direction of its orientation.
Similarly, $m_-(p)$ denotes the limit from the negative side of $\Sigma$. 
Then $m$ is holomorphic away from $\Sigma_{\ell}\cup \Sigma_r$ with jump condition
$m_+(p,n,t)=m_-(p,n,t) v(p,n,t)$ and jump matrix
$$
v(p,n,t)= \left\{
\begin{array}{ll}
\begin{pmatrix}
1 - |R(p)|^2 & - \ol{R(p)} \E^{- 2 t \Phi(p,n/t)} \\[1mm]
R(p) \E^{2 t \Phi(p,n/t)} & 1
\end{pmatrix}, & \quad p \in \Sigma,\\[5mm]
\begin{pmatrix}
0 & - \ol{R(p)} \E^{- 2 t \Phi(p,n/t)} \\[1mm]
R(p) \E^{2 t \Phi(p,n/t)} & 1
\end{pmatrix}, & \quad p \in  \Sigma_r\setminus\Sigma,\\[5mm]
\begin{pmatrix}
\chi(p) \E^{t (\Phi_+(p,n/t)-\Phi_-(p,n/t))} & 1 \\[1mm]
1 & 0 
\end{pmatrix}, & \quad p \in \Sigma_{\ell} \setminus \Sigma,
\end{array}\right.
$$
where $\Sigma$ is the clockwise oriented contour around the spectrum of multiplicity two $I_{\ell} \cap I_r$ (if present). The function $m=(m_1, m_2)$ has positive limiting values as $p \to \infty_\pm$ satisfying $m_1(\infty_\pm) m_2(\infty_\pm)=1$.
In the matrix elements of the jumps, $R(p)$ is the right reflection coefficient at $t=0$ and 
$\chi$ is the limit from the upper sheet $\Pi_U$ of the right transmission coefficient $T(x)$ at $t=0$,
$$
\chi(p) = -\lim_{x \in \Pi_U \to p \in \Sigma_{\ell}} \sqrt{\frac{(x - b_{\ell})^2 - 4a_{\ell}^2}{x^2 - 1}} \ |T(x)|^2.  
$$
The phase function $\Phi$ is given on the closure of the upper sheet by
$$ 
\Phi(p,n/t)=\frac{1}{2} \big(z(p) - z^{-1}(p)\big) + \frac{n}{t}\log z(p)
$$
and continued as an odd function on the lower sheet with a jump on $\Sigma_{\ell} \setminus \Sigma$. 
The expansion of the first component of $m$ as $p \to \infty_+$ yields the precise connection to $a,b$,
$$
m_1(p,n,t) = \prod_{j=n}^\infty 2a(j,t) \Big(1 + \frac{1}{\la}\ \sum_{m=n}^\infty b(m,t)\Big) + O\Big(\frac{1}{\la^{2}}\Big).
$$
If $H$ has eigenvalues, then $m$ is meromorphic away from $\Sigma_{\ell}\cup \Sigma_r$ with pole conditions
$$
\res_{p_j} m(p) = \lim_{p\to p_j} m(p)
\begin{pmatrix} 0 & 0\\ A_j  & 0 \end{pmatrix}, \quad 
\res_{p_j^*} m(p) = \lim_{p\to p_j^*} m(p)
\begin{pmatrix} 0 &  A_j \\ 0 & 0 \end{pmatrix},
$$
where $p_j$ and $p_j^*$ denote the eigenvalue $\la_j$ on the upper and lower sheet and 
$$
A_j= \sqrt{p_j^2-1} \beta_j \E^{2t\Phi(p_j,n,t)}.
$$
Here $\beta_j$, $j=1, \dots, N$, are the right norming constants at time $t=0$.

One tries to find a factorization of the jump matrices in order to transform the initial RHP 
to an equivalent RHP with jump matrices close to constant matrices on contours for large $t$, 
which can be solved explicitly. The asymptotic solution for $a, b$ can then be read off using 
the expansion of $m$ at $\infty_+$. 
The crucial step in the nonlinear stationary phase method \cite{dz} is to reduce
the given RHP to one or more RHPs localized at stationary phase points, which can be analyzed
and controlled individually. However, the steplike case requires an extension of this method 
based on a suitably chosen $g$-function as first introduced in \cite{dvz} which replaces the phase function $\Phi$.
Since the jump contour of the limiting RHP depends on the slow variable $\xi=\frac{n}{t}$, this determines 
a special choice for the $g$-function. The expected asymptotic solution is finite band 
and corresponds to a modified Riemann surface which is "truncated" with respect to the initial Riemann 
surface and moves with $\xi$. So we choose the $g$-function as a sum of Abel integrals 
such that the line $\re g = 0$ passes through the moving end of the truncated Riemann surface and such that
the $g$-function approximates the phase function at infinity up to an additive constant. Then this $g$-function
transforms the jump matrices in a way that allows us to factorize them and to get asymptotically constant 
matrices on contours. In the case of the Toda rarefaction problem with overlapping background spectra, 
the modified Riemann surface corresponds to just one interval, and we describe 
this simple dependence of the asymptotic solution in Sec.~\ref{rareoverlap} in more detail.

\section{Asymptotic solution for non-overlapping background spectra}

\subsection{Toda shock problem} \label{shockno} 
\begin{figure}[ht]
\centering
\includegraphics[width=6cm]{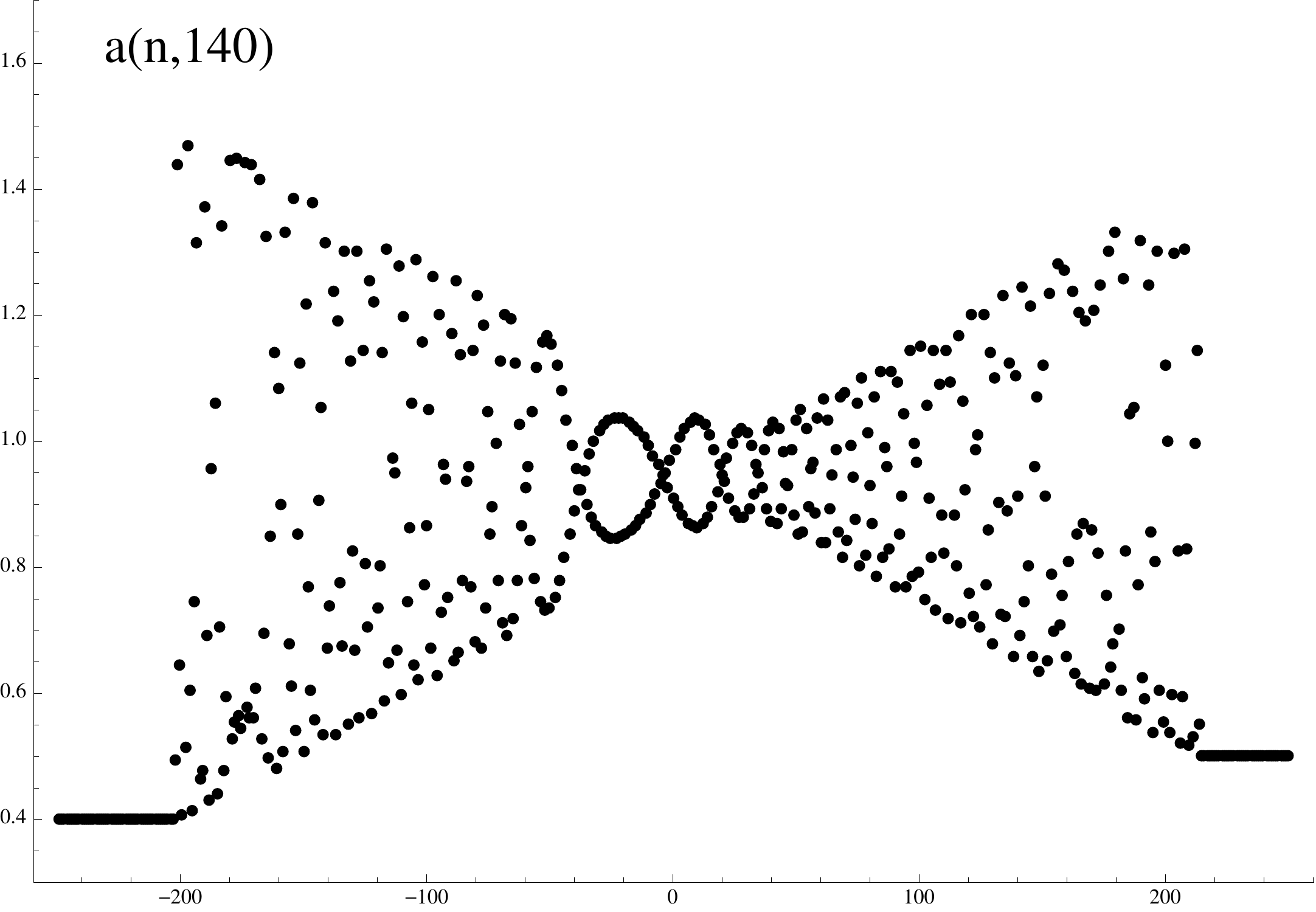}
\hfill
\includegraphics[width=6cm]{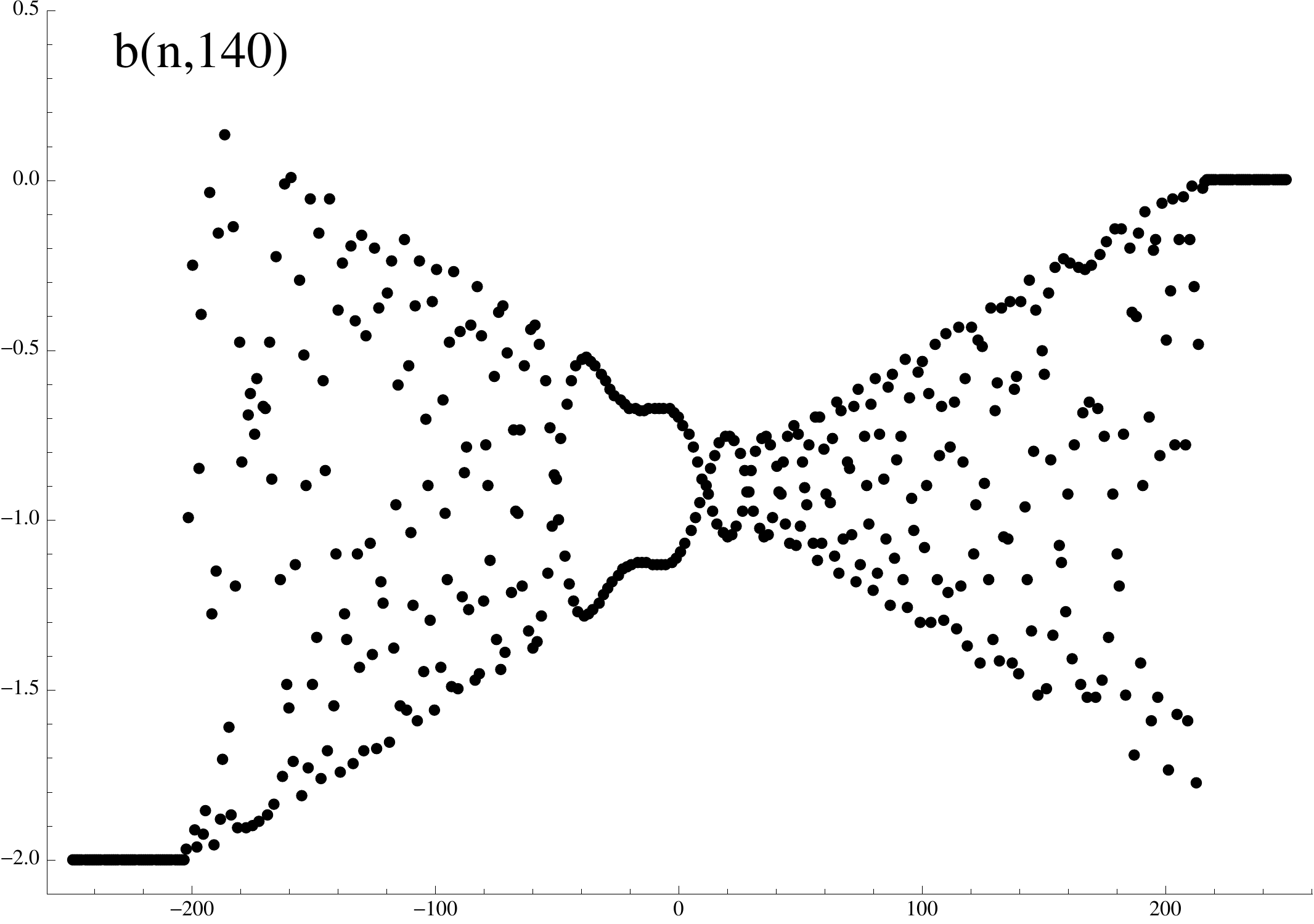}
\caption{{\small Numerically computed solution $a$ and $b$ of the Toda shock problem with $\sigma(H_{\ell})=[-2.8, -1.2]$ and  $\sigma(H_r)=[-1,1]$.}} \label{shocknool}
\end{figure}

Assume that the background spectra are in shock position and do not overlap, $b_{\ell}+2a_{\ell} < -1$. 
Fig.~\ref{shocknool} depicts the numerically computed solution corresponding to the
step $(a(n,0),b(n,0))=(\frac{1}{2}, 0)$ if $n\geq 0$ and $(a(n,0),b(n,0))=(a_{\ell}, b_{\ell})$ if $n< 0$
at $t=140$ with non-overlapping background spectra ($a_{\ell}=0.4$, $b_{\ell}=-2$). 
One can distinuish five principal regions in the half plane $n/t$ divided by 
rays $n_{cr}/t=\xi_{cr}$, where $\xi_{cr}$ is one of four critical points satisfying $\xi_{\ell}< \xi_{\ell}^\prime < \xi_{r}^\prime< \xi_{r}$ (compare also 
Fig.~\ref{elliptic} at a later time $t=270$, where $n_{cr}=\xi_{cr} \times 270$ are plotted
as vertical lines to mark the different regions). 
Analyzing the curve $\re \Phi(\la,\xi)=0$ yields that the critical point $\xi_{r}$ is the solution of 
\be \label{xir}
\int_{\inf I_{\ell}}^{-1} \frac{x+ \xi_{r}}{\sqrt{x^2-1}}dx=0, \quad 
\xi_{r}=\frac{\sqrt{(-b_{\ell}+2a_{\ell})^2 -1}}{\log (-b_{\ell}+2a_{\ell} + \sqrt{(-b_{\ell}+2a_{\ell})^2-1})},
\ee
the remaining points $\xi_{cr}$ are given in \cite{emt14}.
In the region $\xi > \xi_r$, the solution is asymptotically close to the constant right background solution 
$(\frac{1}{2}, 0)$, and in the domain $\xi<\xi_{\ell}$, it is close to the left background $(a_{\ell}, b_{\ell})$.

\begin{figure}[ht]
\centering
\includegraphics[width=12cm]{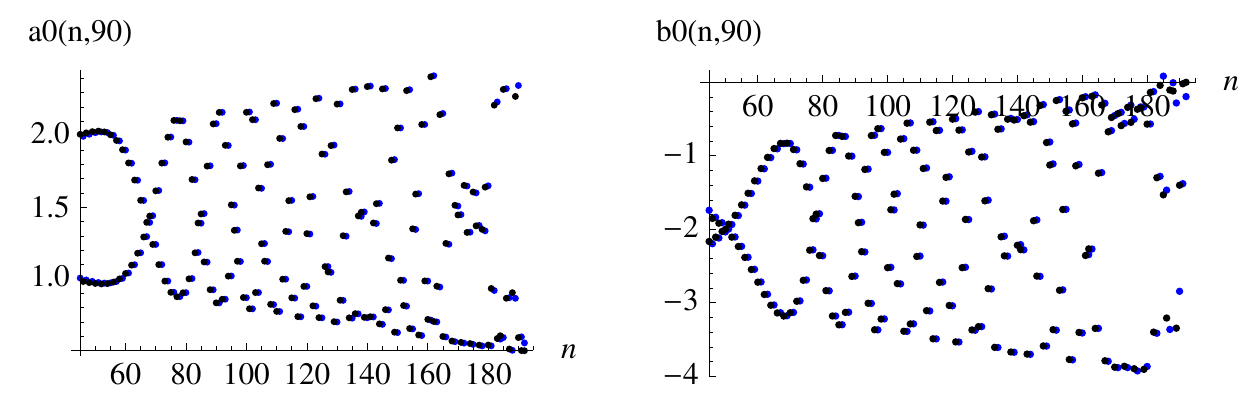}
\caption{{\small Comparison between the solution (black) and the asymptotic 
formula (blue) in the region $\xi_r^\prime<\xi<\xi_r$.}}\label{fig2}
\end{figure}

\begin{figure}[ht]
\centering
\includegraphics[width=13cm]{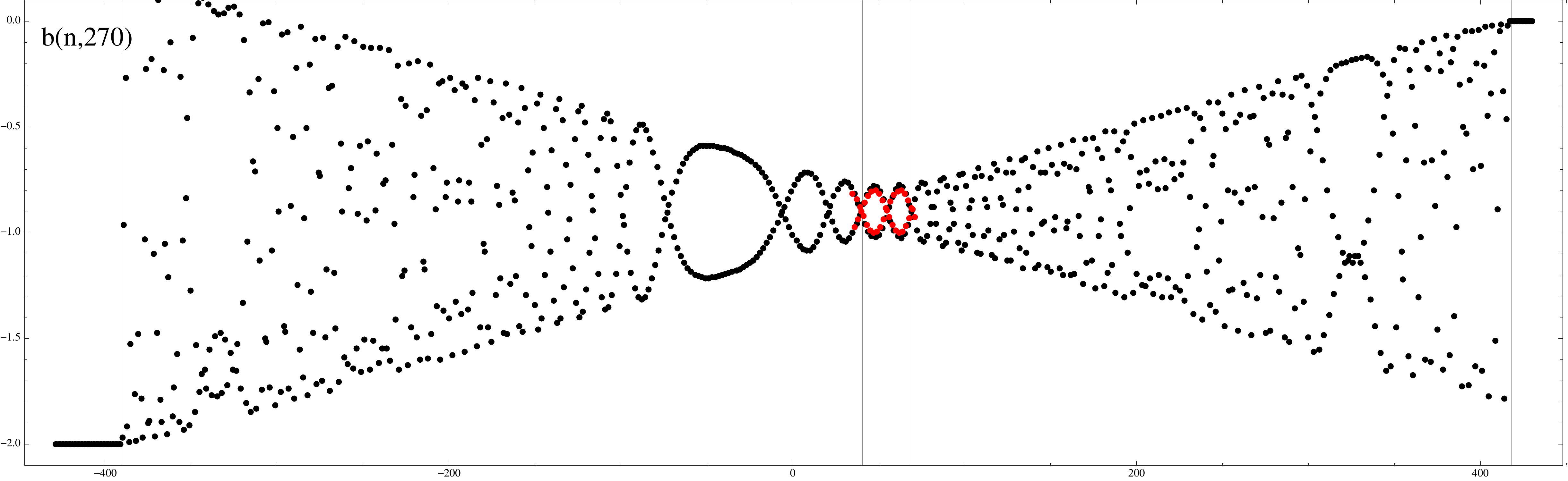}
\caption{{\small Comparison between the $b$ coefficient (black) of Fig.~\ref{shocknool} at time $t=270$  
and the asymptotic formula (red) in the region $\xi_{\ell}^\prime < \xi < \xi_r^\prime$. 
}} 
\label{elliptic}
\end{figure}

In the domain $\xi_r^\prime< \xi <\xi_r$, we find 
a monotonic smooth function $\gamma(\xi) \in I_{\ell}$ such that
 $\gamma(\xi_r^\prime)=\sup I_{\ell}$, $\gamma(\xi_r)=\inf I_{\ell}$.
When the parameter $\xi$ starts to decay from the point $\xi_r$, the point $\gamma(\xi)$ ``opens'' a band 
$[\inf I_{\ell},\gamma(\xi)]=I_{\ell}(\xi)$ (the Whitham zone). The intervals $I_{\ell}(\xi)$ and $I_r$ 
can be treated as the bands of a (slowly modulated) finite band solution of the Toda lattice, which turns out to give the leading asymptotic term of our solution with respect to large $t$. This finite band solution is modulated by the gradual lengthening of the lower band and defined uniquely by its initial divisor. We compute this divisor precisely via the values of the right transmission coefficient on the interval $I_{\ell}(\xi)$. Thus, in a vicinity of any ray $n/t=\xi$ the solution of \eqref{tl}--\eqref{ini1} is asymptotically two band.
This asymptotic term also can be treated as a function of $n$, $t$, and $\frac{n}{t}$ in the whole domain $t(\xi_r^\prime+\varepsilon)<n<t(\xi_r-\varepsilon)$. 
To obtain analytical formulas, one shows that for any  $\xi \in (\xi_r^\prime, \xi_r)$ 
there exist $\gamma(\xi) \in I_{\ell}$ and $\mu(\xi) \in (\gamma(\xi), -1)$ satisfying
$$
2a_{\ell}-b_{\ell}+ \gamma(\xi) + 2 \mu(\xi)=-2\xi, \quad 
\int_{\gamma(\xi)}^{-1}\frac{\big(\la - \mu(\xi)\big)\big(\la - \gamma(\xi)\big)}{P(\la, \gamma)}d\la = 0,
$$
where
$$
P(\la,\gamma)=- \sqrt{(\la^2-1)(\la-b_{\ell}+2a_{\ell})(\la - \gamma(\xi))}.
$$
Then the $g$-function with the desired properties to transform the jump matrices is given by (see \cite[Sec.\ 3]{emt14})
$$
g(p, \xi) =  \int_{1}^p
\frac{\big(\la - \mu(\xi)\big)\big(\la-\gamma(\xi)\big)}{P(\la,\gamma)} d \la.
$$
Associated with the square root $P(\la,\gamma)$ is the "truncated" Riemann surface $\M(\xi)$.
Let $\zeta$ be the normalized holomorphic Abel differential (which depends on $\xi$) on $\M(\xi)$.
The divisor $\rho(\xi)$ of the two band solution under consideration is uniquely defined by the following 
Jacobi inversion problem 
$$
\int_{\inf I_{\ell}}^{\rho(\xi)} \zeta = \int_{\inf I_{\ell}}^{\gamma(\xi)} \zeta - \frac{\I}{\pi}\int_{\inf I_{\ell}}^{\gamma(\xi)}\log|\chi| \zeta + \int_{\infty_-}^{\infty_+}\zeta.
$$
Introduce the function 
$$
 \underline{z}(n,t)=\int_{\inf I_{\ell}}^{\infty_+} \zeta - \int_{\inf I_{\ell}}^{\rho(\xi)} \zeta - n \int_{\infty_-}^{\infty_+}\zeta +\frac{t}{\pi\I}\int_{\inf I_{\ell}}^{\gamma(\xi)} \Omega_0 - \Xi(\xi),
$$
where $\Xi(\xi)$ is the Riemann constant and $\Omega_0$ is the normalized Abel differential of the second kind on $\M(\xi)$ with second order poles at $\infty_-$ and $\infty_+$. Let
\[
\theta(v)=\sum_{m\in\Z} \exp\big(\pi\I m^2\tau + 2\pi\I m v\big)
\]
be the Riemann theta function of the surface $\M(\xi)$ and set
\begin{align} \label{twoband}
	\begin{split}
b_q(n,t,\xi) & =\tilde b +\frac{1}{\Gamma}\,\frac{\pa}{\pa w}\log\left(\frac{
\theta\big(\underline z(n-1,t) + w\big)}{\theta\big(\underline z(n,t) + w\big)}\right)\Big|_{w=0}, \\ 
a^2_q(n,t,\xi)& =\tilde a^2\,\frac{\theta(\underline{z}(n-1,t))\theta(\underline{z}(n+1,t))}{\theta^2(\underline{z}(n,t))},
\end{split}
\end{align}
which describe a classical two band Toda lattice motion corresponding 
to the bands $I_{\ell}(\xi)$ and $I_r$ with initial divisor $\rho(\xi)$ (compare \cite[Sec.\ 9]{tjac}).
Here $\tilde a$, $\tilde b$ are the averages and $\Gamma=\int_{\mathfrak a} \zeta$. 
Then for $\xi \in (\xi_r^\prime, \xi_r)$ in the vicinity of any ray $n=\xi t$, the solution 
has the asymptotic behavior as $t \to +\infty$
\[
a^2(n,t)=a^2_q(n,t,\xi) +o(1), \quad b(n,t)=b_q(n,t,\xi) +o(1).
\]
A numerical comparison between the solution and the asymptotic formula %in this region 
is given in Fig.~\ref{fig2} (due to \cite{emt14}). There we computed the 
two band Toda solution (blue) precisely via \eqref{twoband} for the pure step initial data 
and plotted it against the numerical solution (black) with the same initial data. 

In the gap domain $\xi_{\ell}^\prime < \xi < \xi_r^\prime$, the asymptotic of the solution 
of \eqref{tl}--\eqref{ini1} is described by a two band Toda lattice solution connected with one and the same 
intervals $I_{\ell}$ and $I_r$ and the initial divisor $\rho$ (or shift of the phase) defined by 
$$
\int_{\inf I_{\ell}}^{\rho} \zeta = \int_{\inf I_{\ell}}^{\sup I_{\ell}} \zeta 
- \frac{\I}{2 \pi}\int_{\Sigma_{\ell}}\log|\chi| \zeta
+ \int_{\infty_-}^{\infty_+}\zeta
$$
does not depend on the slow variable $\xi$. Here $\zeta$ is the normalized holomorphic Abel 
differential on the initial Riemann surface associated with $P(\la)$ (cf. \eqref{rs}). 
Then the solution of \eqref{tl}--\eqref{ini1} is asymptotically close to 
$a_q(n,t), b_q(n,t)$ constructed as above, but independent of $\xi$,  
as $t\to \infty$ uniformly in $n/t \in [\xi_{\ell}^\prime +\varepsilon, \xi_r^\prime -\varepsilon]$. 
For the comparison with the numerical solution
in Fig.~\ref{elliptic}, we expressed $a_q, b_q$ in terms of Jacobi's elliptic functions 
(cf. \cite[Sec. 9.3]{tjac}). 
Note that if solitons were present in the gap between the background spectra, new 
two band solutions associated with $I_{\ell}$ and $I_r$ would appear to the left and to the right of each soliton, 
differing by a phase shift (see \cite{emt14}). If the background spectra are of equal length, $a_{\ell}=\frac{1}{2}$, 
the solution $a_q, b_q$ is periodic (Fig.~\ref{shocknools} and \cite{vdo}).

In the second Whitham zone $\xi_{\ell}<\xi < \xi_{\ell}^\prime$, there is a monotonic smooth function 
$\gamma_{\ell}(\xi) \in I_r$ with $\gamma_{\ell}(\xi_{\ell})=\sup I_r$, $\gamma_{\ell}(\xi_{\ell}^\prime)=\inf I_r$. 
The modulated finite band asymptotic here is local along the ray and defined by the intervals $I_{\ell}$,  
$[\gamma_{\ell}(\xi), \sup I_r]$, and an initial divisor.

\subsection{Toda rarefaction problem} \label{ss:no}
Consider the rarefaction problem without overlap of the background spectra, so let $\sup I_r=1 < \inf I_{\ell}$. 
\begin{figure}[ht]
\centering
\includegraphics[width=5cm]{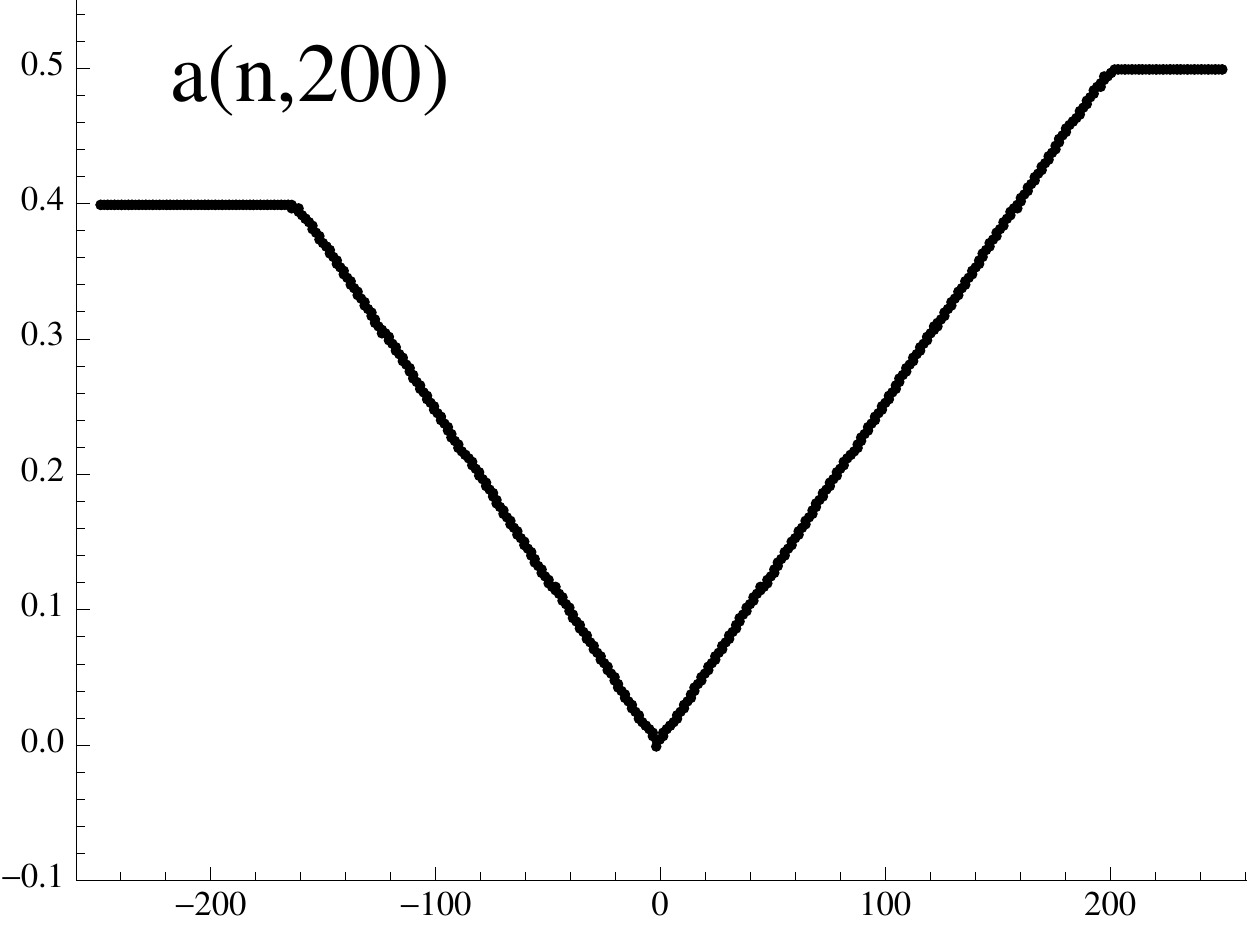}
\hspace{1cm}
\includegraphics[width=5cm]{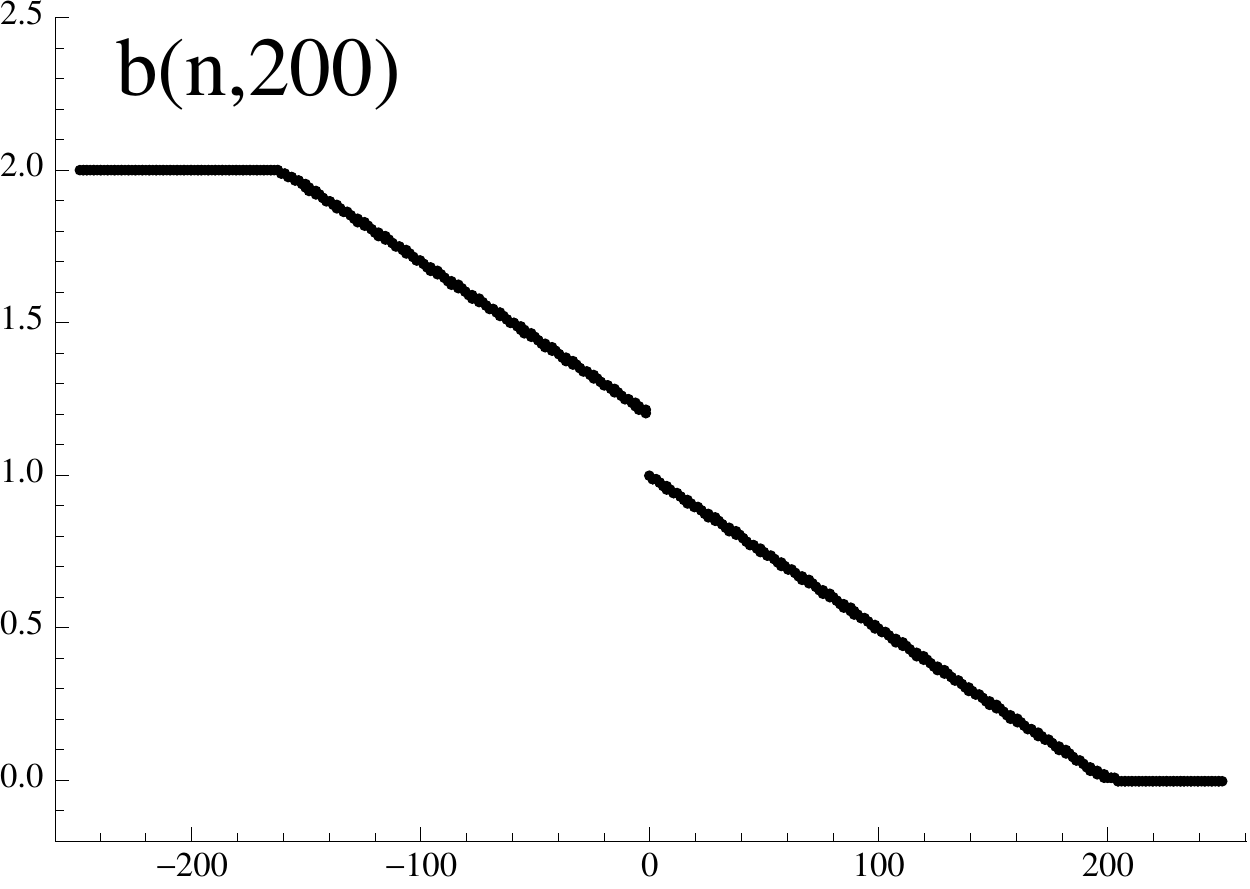}
\caption{{\small Toda rarefaction problem with non-overlapping background spectra $\sigma(H_{\ell})=[1.2, 2.8]$,   $\sigma(H_r)=[-1,1]$; $a_{\ell}=0.4$, $b_{\ell}=2$, $a_r=\frac{1}{2}$, $b_r=0$.  \label{rarenool} }}
\end{figure}
As illustrated in Fig.~\ref{rarenool}, the $n/t$ plane splits into four main regions.
In the domain $\xi>1$, the solution is asymptotically close to the constant right background solution $(\frac{1}{2},0)$
and in the domain $\xi < -2a_{\ell}$, it is close to the left background solution. For 
$\xi \in (0,1)$, the solution is asymptotically close to  
\be \label{slopesol1}
a^2(n,t)=\left(\frac{n}{2t}\right)^2 + o(1), \quad 
b(n,t)= 1 - \frac{n}{t} + o(1).
\ee
In the domain $\xi \in (-2a_{\ell}, 0)$,
\be \label{slopesol2}
a^2(n,t)=\left(\frac{n}{2t}\right)^2 + o(1), \quad 
b(n,t)= b_{\ell}- 2a_{\ell}- \frac{n}{t} + o(1).
\ee
The proof of these asymptotics will be provided in a forthcoming paper.

\section{Asymptotic solution for overlapping background spectra}

In this section we assume that the background spectra $I_{\ell}$ and $I_r$ overlap, which means that 
the Jacobi operator $H(t)$ has a nonempty spectrum of multiplicity two. The asymptotic solution 
on the region corresponding to this spectrum  
is given by the constant solution associated with the interval $[b_{\ell} - 2 a_{\ell}, 1]$ plus a
dispersive tail which decays like $t^{-1/2}$, 
\begin{align} \label{constantsol}
\begin{split}
a^2(n,t) &=\big((1-b_{\ell}+2a_{\ell})/4\big)^2 + \mbox{oscillation}, \\
b(n,t) &= (1 + b_{\ell})/2 - a_{\ell} + \mbox{oscillation}.
\end{split}
\end{align}
The oscillatory tail is in agreement with the decaying case \cite{KTb}. 
In particular, the asymptotic \eqref{constantsol} holds true for any type of overlap of $I_{\ell}$ and $I_r$.

\subsection{Toda shock problem with dispersive tail}
\begin{figure}[ht]
\centering
\includegraphics[width=6.2cm]{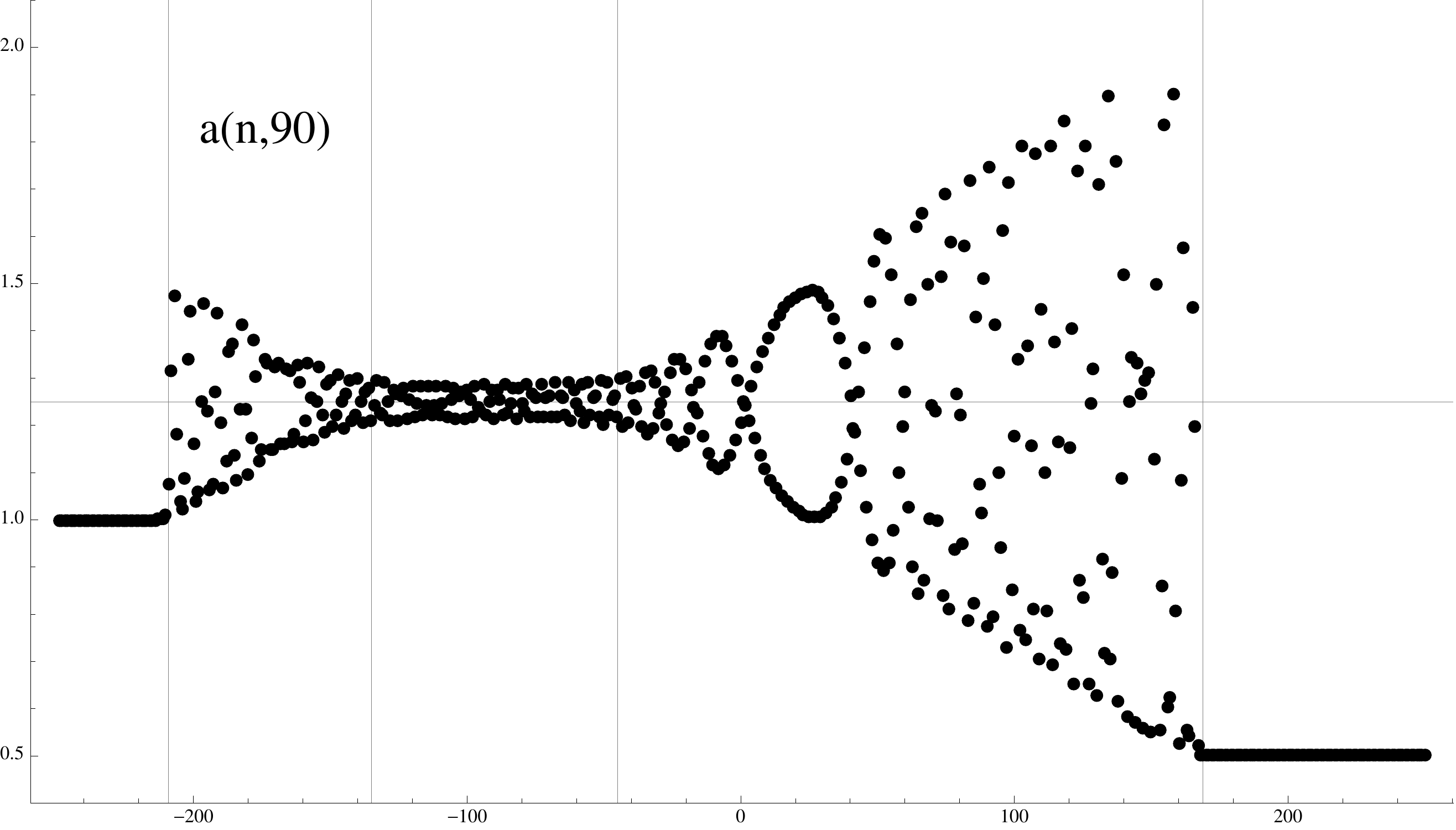}
\hfill
\includegraphics[width=6.2cm]{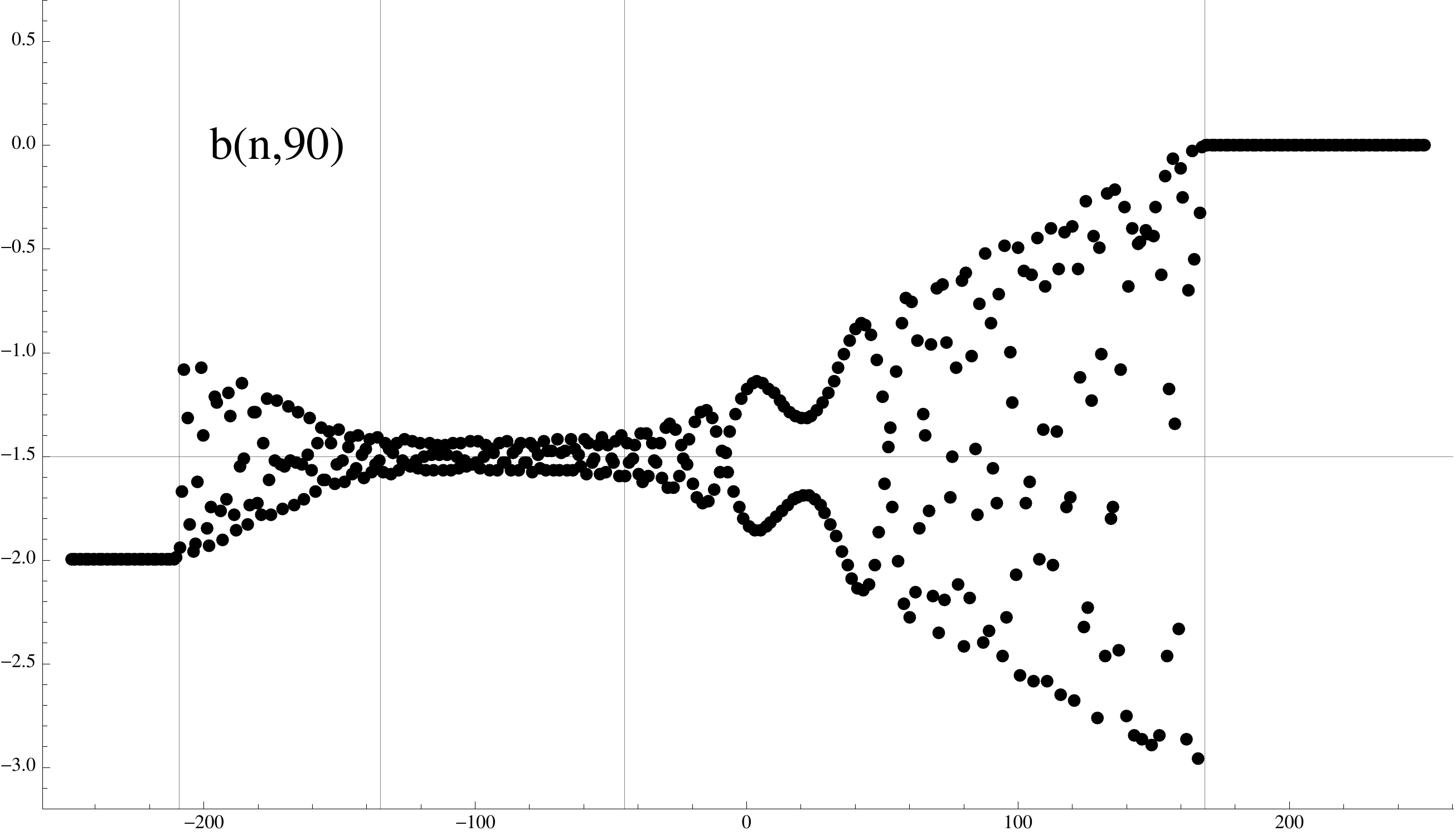}
\caption{{\small Toda shock with overlapping background spectra $\sigma(H_{\ell})=[-4, 0]$ and  $\sigma(H_r)=[-1,1]$;
$a_{\ell}=1$, $b_{\ell}=-2$.
}} \label{shockol}
\end{figure}

Let the background spectra overlap such that $\inf I_{\ell} < -1 < \sup I_{\ell} < 1$. Then 
the $n/t$ plane splits into five different regions with boundary points
$$
\xi_{\ell}< \frac{1-b_{\ell}-6 a_{\ell}}{2}< \frac{b_{\ell}-2a_{\ell} +3}{2} < \xi_r,
$$
where $\xi_r$ and $\xi_{\ell}$ are the same as in Sec.~\ref{shockno}. 
The boundary values are plotted in Fig.~\ref{shockol} as vertical lines. On the right 
Whitham zone, $(b_{\ell}-2a_{\ell} +3)/2 < \xi < \xi_r$, the 
function $\gamma(\xi) \in [\inf I_{\ell},-1]$ exists and the asymptotic solution is the two band 
Toda lattice solution $a_q^2, b_q$ with initial divisor 
$\rho(\xi)$ and bands $[\inf I_{\ell},\gamma(\xi)]$ and $I_r$ of Sec.~\ref{shockno}. 
In the middle region corresponding to the spectrum of multiplicity two, 
the asymptotic solution is given by \eqref{constantsol} which we illustrate 
by plotting the constant term of \eqref{constantsol} as a horizontal line. This line is 
at $1.25$ for $a(n,90)$ and at $-1.5$ for $b(n,90)$. 
On the left Whitham zone, there exists a monotonic smooth function 
$\gamma_{\ell}(\xi) \in [\max I_{\ell}, 1]$ 
and the modulated two band Toda lattice asymptotic here is defined by the intervals $I_{\ell}$ and   
$[\gamma_{\ell}(\xi), 1]$, and an initial divisor as before. 

\subsection{Toda rarefaction problem with dispersive tail} \label{rareoverlap}
\begin{figure}[ht]
\centering
\includegraphics[width=6cm]{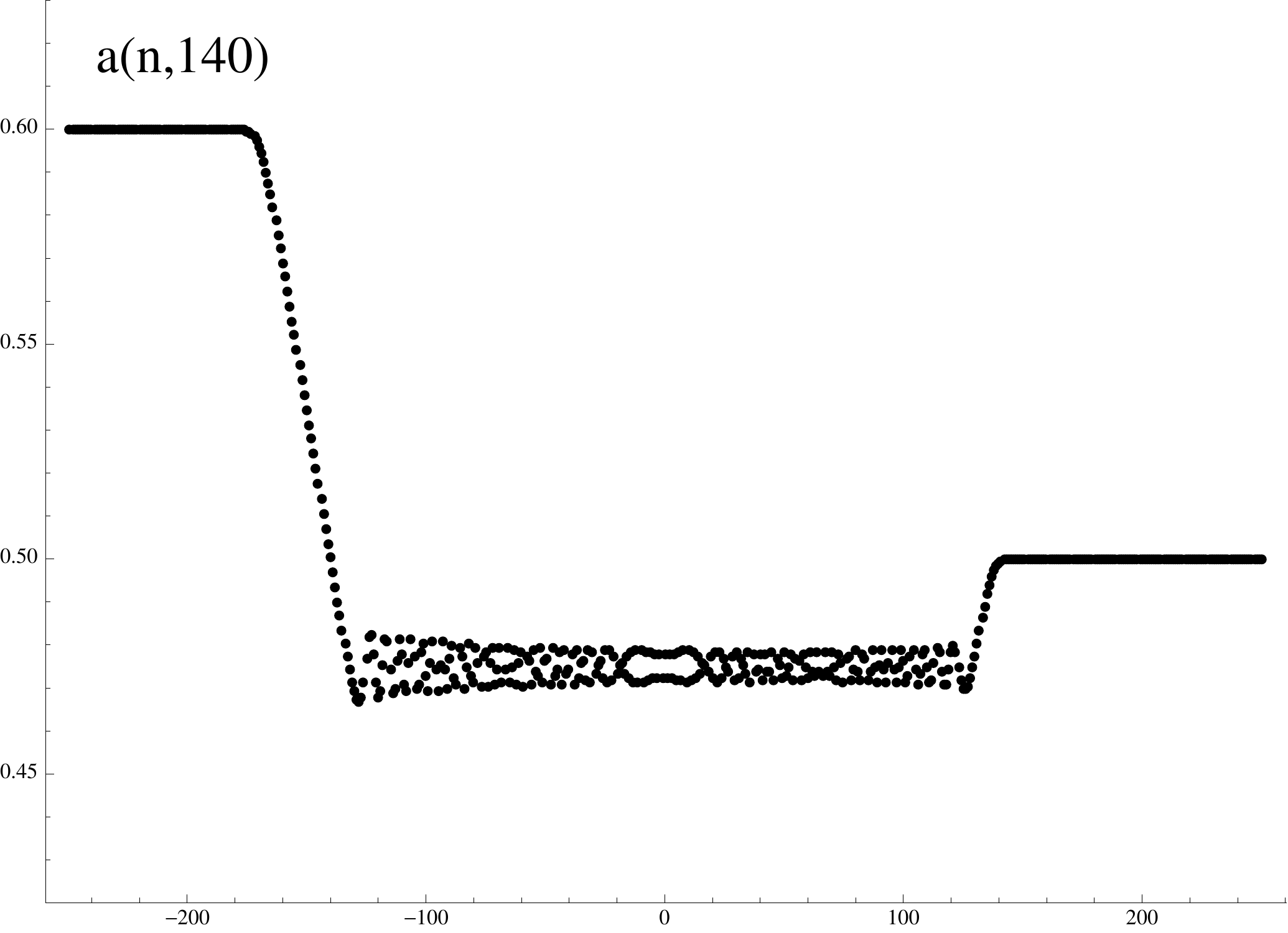}
\hfill
\includegraphics[width=6cm]{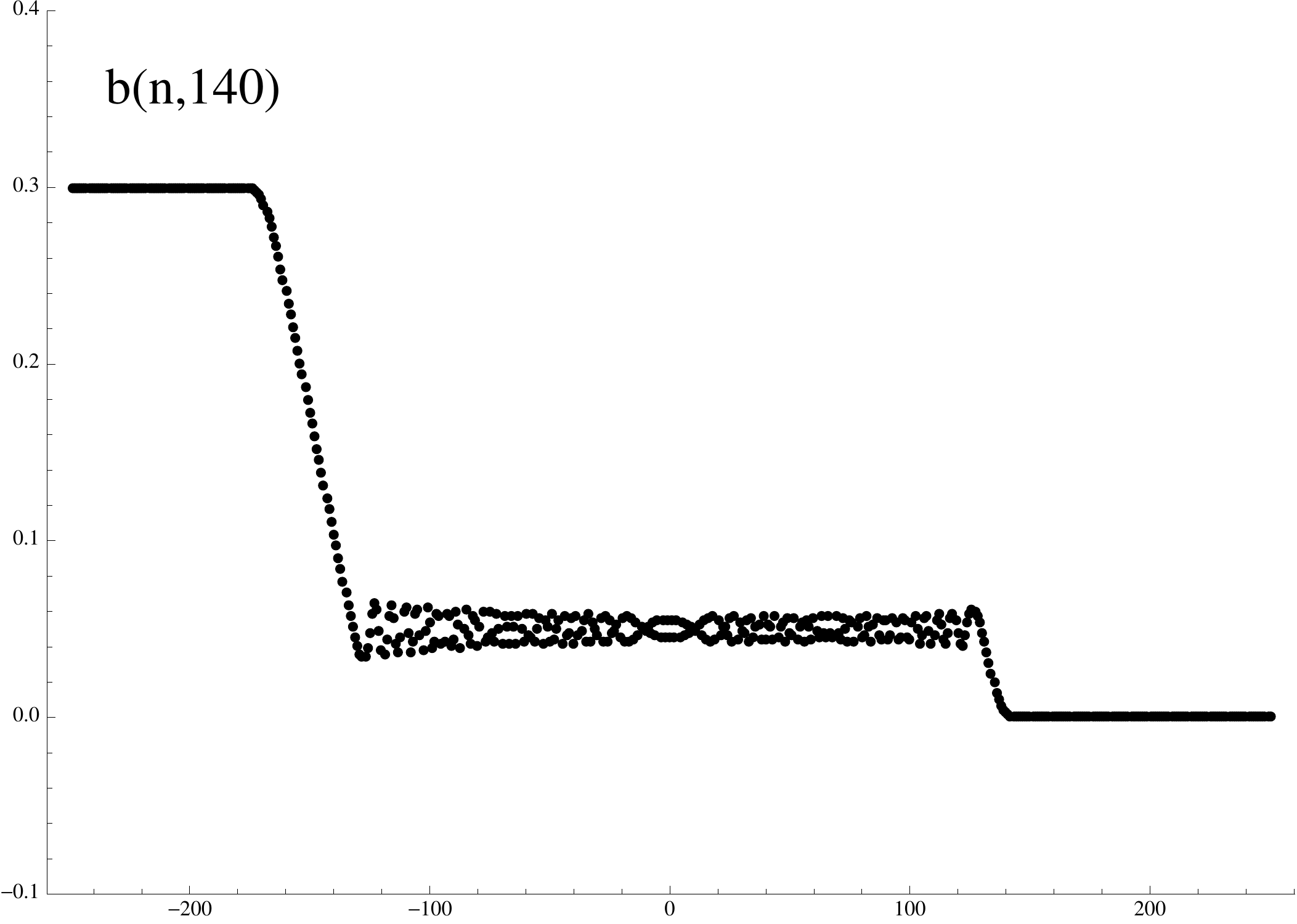}
\caption{{\small Toda rarefaction with overlapping background spectra $\sigma(H_{\ell})=[-0.9, 1.5]$ and $\sigma(H_r)=[-1,1]$;  $a_{\ell}=0.6$, $b_{\ell}=0.3$.}} \label{rareol}
\end{figure}

Let the background spectra overlap such that $-1 < \inf I_{\ell} < 1 < \sup I_{\ell}$. 
Then the $n/t$ plane splits into five regions with boundary points
$$
-2a_{\ell} < \frac{b_{\ell}-2a_{\ell}-1}{2} < \frac{1-b_{\ell}+2a_{\ell}}{2} < 1.
$$
In Fig.~\ref{rareol}, the slopes correspond to the spectra of multiplicity one, while the oscillating part 
is due to the spectrum of multiplicity two. 
For this problem the underlying Riemann surface of the 
$g$-function consists of just one interval, which continuously transforms from $I_r$ to $I_{\ell}$ 
as $\xi$ decreases from $+\infty$ to $-\infty$ and determines the leading term of the asymptotic solution, so
let us describe this situation in more detail. The difficulty is to find the correct $g$-function which replaces the 
phase function $\Phi$ in the RHP. The $g$-function differs for each region with matching definitions 
at the respective boundary points. Let $\eta(\xi)$ be the point where the curve $\re g=0$ crosses the real axis. 
As $\xi$ decreases, $\eta(\xi)$ increases from $-\infty$ to $+\infty$.
For $\xi > 1$, the Riemann surface corresponds to $I_r$ and the solution is 
asymptotically close to $(\frac{1}{2},0)$.
When $\xi$ starts to decay from $1$, the point $\eta(\xi)=1-2\xi$ truncates $I_r$ to $[\eta(\xi), 1]$
and the solution is asymptotically close to \eqref{slopesol1}.
When $\xi$ passes the second boundary point $(1-b_{\ell}+2a_{\ell})/2$, the point $\eta(\xi)$ passes $\inf I_{\ell}$ and 
the Riemann surface corresponds to $[\inf I_{\ell}, 1]$. The asymptotic solution is given by the constant solution with dispersive tail \eqref{constantsol}.
When $\xi$ is at the third boundary point, $\eta(\xi)$ crosses $1$, and the interval starts to enlarge, 
$[\inf I_{\ell}, \eta(\xi)]$, with $\eta(\xi)=b_{\ell}-2a_{\ell}-2\xi$. The solution is asymptotically close to 
\eqref{slopesol2}.
When $\xi$ passes $-2a_{\ell}$, $\eta(\xi)$ is at $\sup I_{\ell}$, and the Riemann surface corresponds to $I_{\ell}$,
so the asymptotic solution is $(a_{\ell},b_{\ell})$.

\subsection{Mixed cases of embedded background spectra} \label{mixed} 
There are two cases to consider, either $I_r$ is embedded in $I_{\ell}$ or $I_{\ell}$ is a subset of $I_r$.

\begin{figure}[ht]
\centering
\includegraphics[width=6cm]{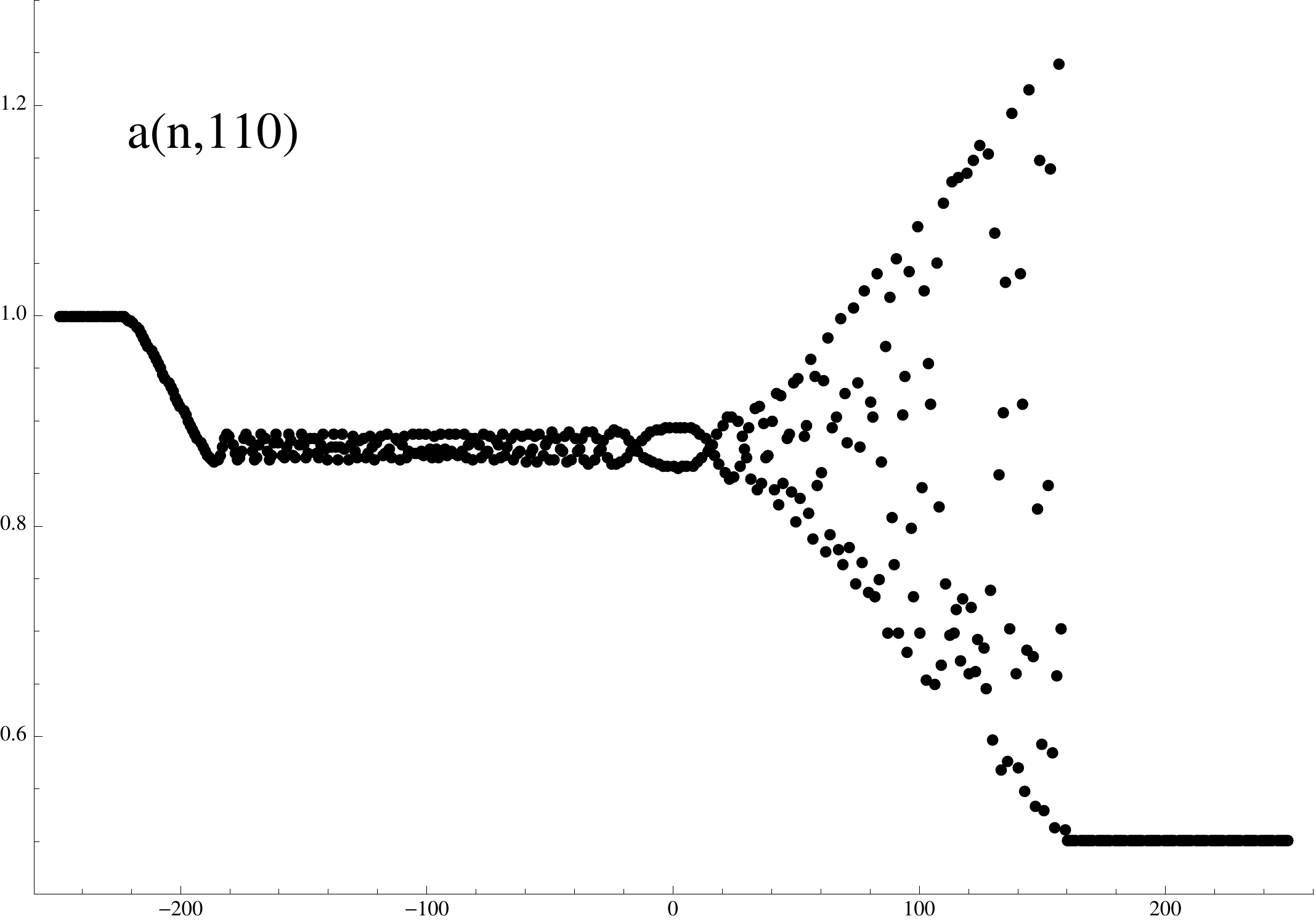}
\hfill
\includegraphics[width=6cm]{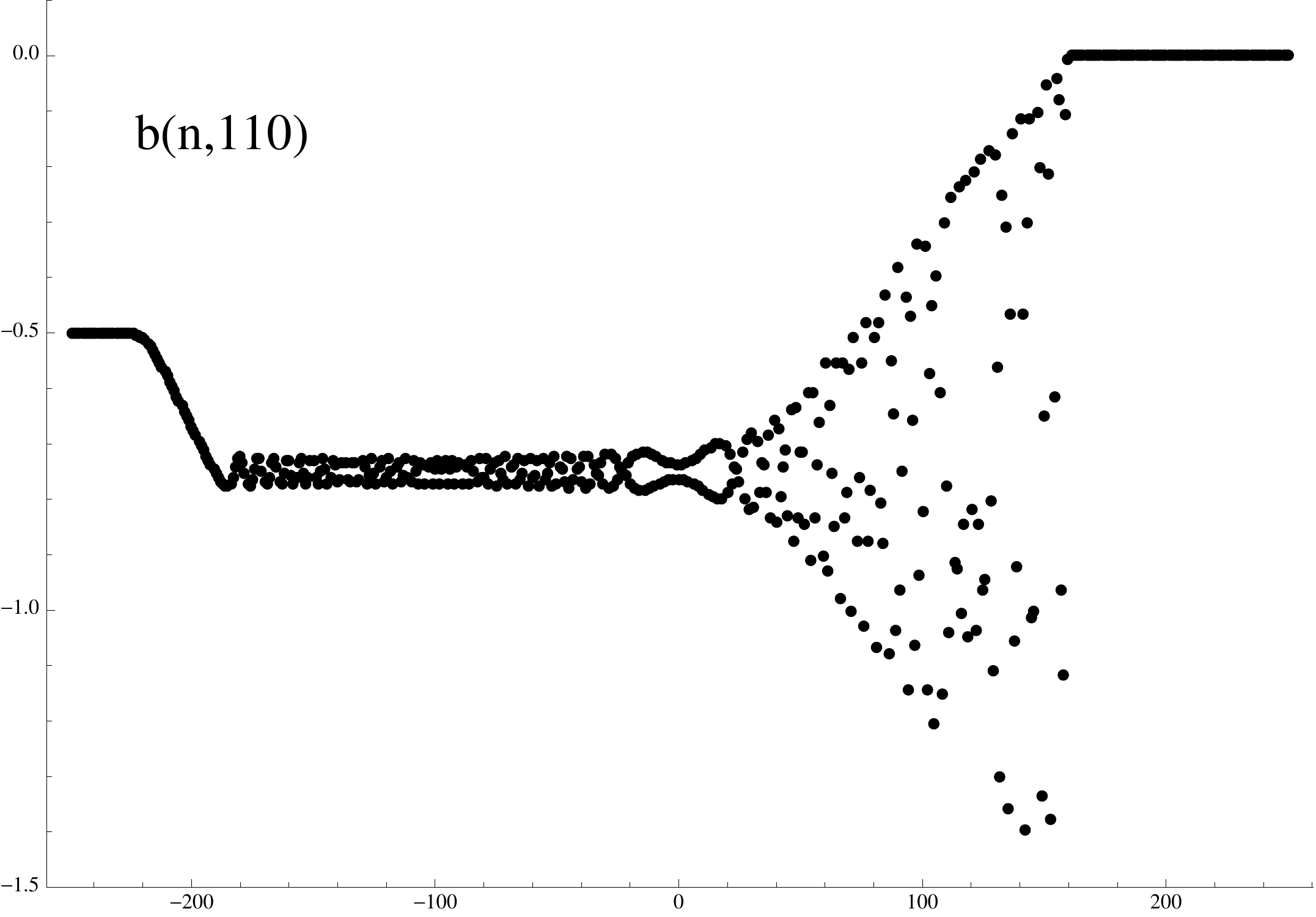}
\caption{{\small Mixed case with background spectrum 
$\sigma(H_r)=[-1,1]$ embedded in $\sigma(H_{\ell})=[-2.5, 1.5]$; $a_{\ell}=1$, $b_{\ell}=-\frac{1}{2}$.
}} \label{shockem}
\end{figure}

\noindent {\bf Case 1:} Let   $I_r\subset I_{\ell}$ (compare Fig.~\ref{shockem}).
The boundary points of the different regions are given by
$$
-2a_{\ell}< \frac{b_{\ell}-2a_{\ell} -1 }{2} < \frac{b_{\ell}-2a_{\ell} +3}{2} < \xi_r,
$$
with $\xi_r$ as in \eqref{xir}.
The two band solution of the Whitham zone is defined by the intervals $[\inf I_{\ell}, \gamma(\xi)]$
and $[-1, \sup I_{\ell}]$ with divisor $\rho_e(\xi)$, 
where the normalized holomorphic Abel differential used in the definition of $\rho_e(\xi)$ involves
the square root $- \sqrt{(\la+1)((\la-b_{\ell})^2-4a_{\ell}^2)(\la-\gamma)}$.
In the middle region, the asymptotic is given by the constant solution \eqref{constantsol} with dispersive tail 
and in the region $-2a_{\ell} < \xi < \frac{b_{\ell}-2a_{\ell} -1 }{2}$, it is given by \eqref{slopesol2}.
For $\xi < -2a_{\ell}$ and $\xi >  \xi_r$, the particles are close to the unperturbed lattice. 

\begin{figure}[ht]
\centering
\includegraphics[width=6cm]{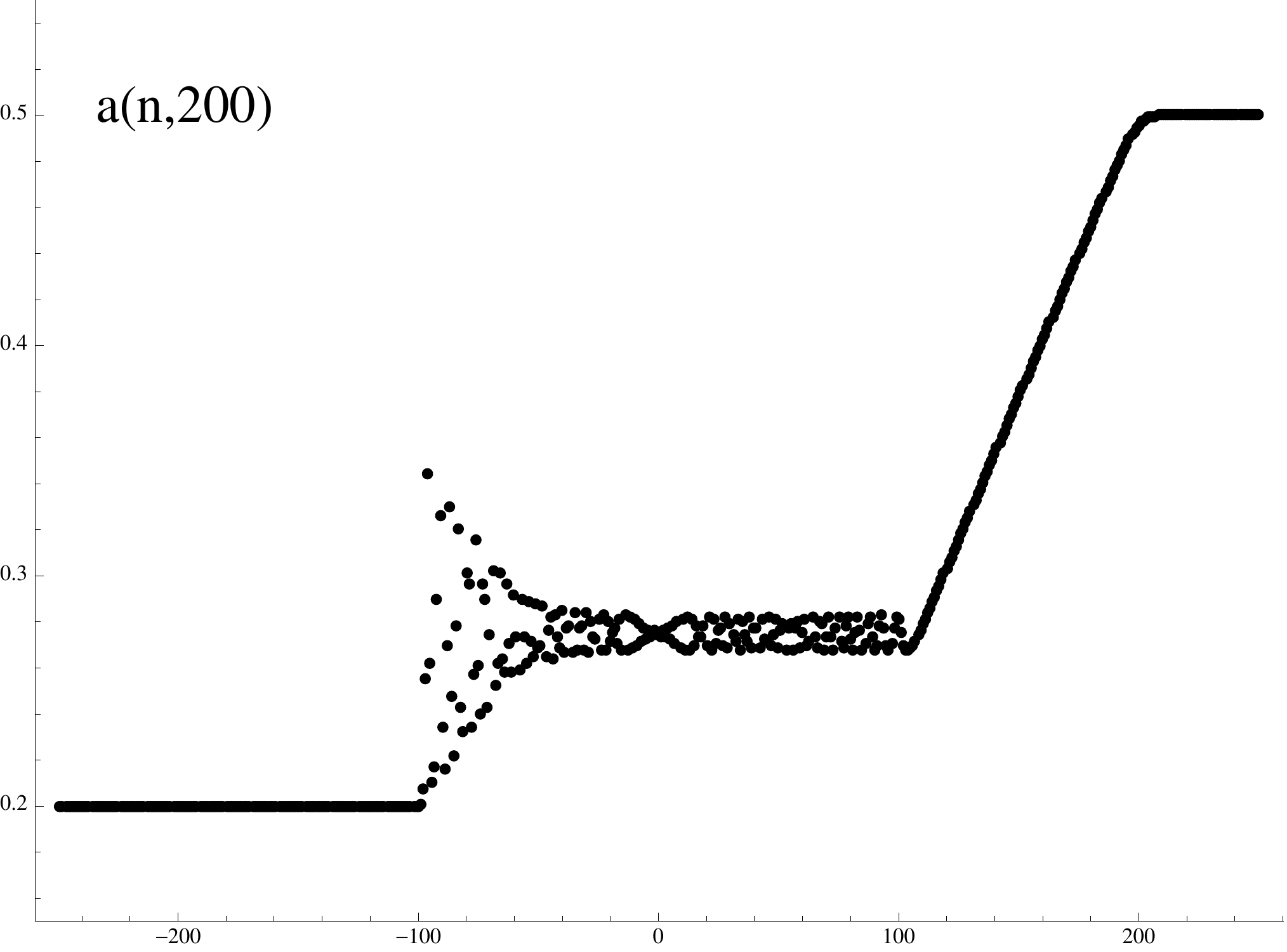}
\hfill
\includegraphics[width=6cm]{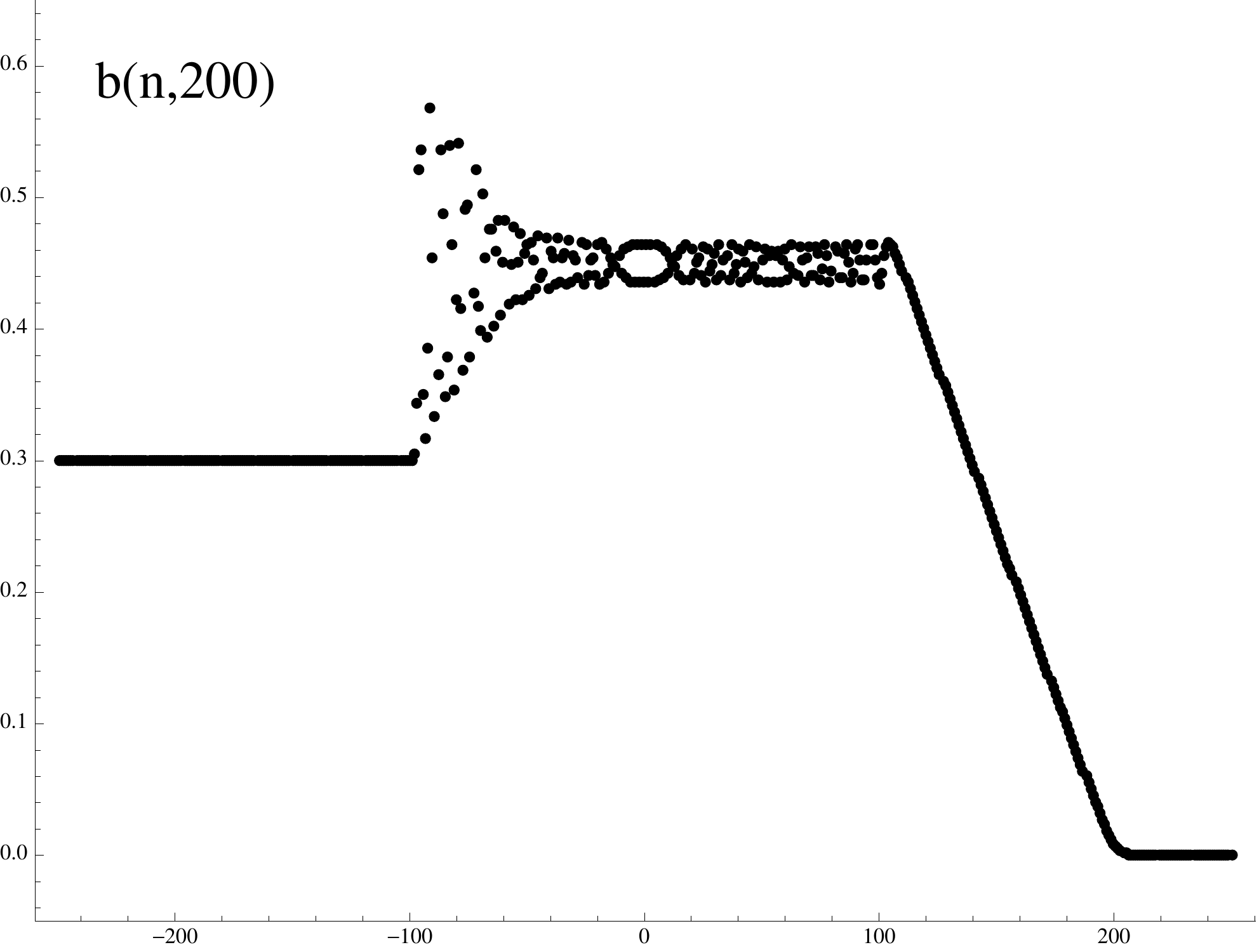}
\caption{{\small Mixed case with background spectrum $\sigma(H_{\ell})=[-0.1, 0.7]$ embedded in $\sigma(H_r)=[-1,1]$; $a_{\ell}=0.2$, $b_{\ell}=0.3$.}} \label{rareem}
\end{figure}

{\bf Case 2}: Let $I_{\ell} \subset I_r$. The boundary points of the different regions in Fig.~\ref{rareem} are given by 
$$ 
\xi_{cr} < \frac{1-b_{\ell}}{2}-3a_{\ell} < \frac{1-b_{\ell}}{2} + a_{\ell} < 1,
$$ 
where $\xi_{cr}$ is the solution of 
$$
\int_{\sup I_{\ell}}^1\frac{x-b_{\ell}+ \xi_{cr}}{
\sqrt{((x -b_{\ell})^2 -4a_{\ell}^2)}}dx=0.
$$
For $\xi > 1$, the solution is asymptotically close to the right background solution, the interval corresponding to the
underlying Riemann surface of the $g$-function is $I_r$. When $\eta(\xi)$ passes $-1$, $I_r$ is shortened to $[\eta(\xi), 1]$ 
and the asymptotic solution is \eqref{slopesol1} until $\eta(\xi)$ passes $\inf I_{\ell}$. Then the Riemann surface corresponds to $[\inf I_{\ell}, 1]$ with asymptotic solution \eqref{constantsol} until $\eta(\xi)$ crosses $\sup I_{\ell}$, 
at which point a gap opens and the surface corresponds to $I_{\ell} \cup {[\eta(\xi),1]}$. 
Hence for $\xi \in (\xi_{cr}, \frac{1-b_{\ell}}{2} -3a_{\ell})$, the asymptotic solution is the two band Toda lattice
solution associated with the bands $I_{\ell}$, $[\eta(\xi),1]$, and an initial divisor. 
Here $\eta(\xi)= 1 + 2b_{\ell}-2\xi - 2\mu(\xi)$ and the function 
$\mu(\xi) \in (\sup I_{\ell},\eta(\xi))$ is the solution of  
$$
\int_{\sup I_{\ell}}^{\eta(\xi)} \frac{(x-\mu(\xi))\sqrt{x-\eta(\xi)}}{
\sqrt{((x - b_{\ell})^2 -4a_{\ell}^2)(x-1)}}dx=0. 
$$
When $\eta(\xi)$ crosses $1$ (as $\xi$ crosses $\xi_{cr}$),  we are left with 
$I_{\ell}$.

\bigskip
\noindent {\bf Acknowledgment.} The author is indebted to Iryna Egorova and the referee for constructive remarks. 
This work was supported by the Austrian Science Fund (FWF) under Grant No.\ V120.

\end{document}